\documentclass[prl,twocolumn,showpacs]{revtex4}

\usepackage{graphicx}
\usepackage{dcolumn}
\usepackage{amsmath}

\begin{document}

\title{Dynamic Resonance of Light in Fabry-Perot Cavities} 

\author{M. Rakhmanov} \author{R.L. Savage, Jr.}
\altaffiliation{LIGO Hanford Observatory, P.O. Box 1970, M/S S9-02, 
Richland, WA 99352}
\author{D.H. Reitze} \author{D.B. Tanner}
\affiliation{Physics Department, P.O. Box 118440, 
University of Florida, Gainesville, FL 32611}

\date{January, 2001}

\begin{abstract}
The dynamics of light in Fabry-Perot cavities with varying length 
and input laser frequency are analyzed and the exact condition for
resonance is derived. This dynamic resonance depends on the light
transit time in the cavity and the Doppler effect due to the mirror
motions. The response of the cavity to length variations is very
different from its response to laser frequency variations. If the
frequency of these variations is equal to multiples of the cavity free
spectral range, the response to length is maximized while the response
to the laser frequency is zero. Implications of these results for the
detection of gravitational waves using kilometer-scale Fabry-Perot
cavities are discussed.
\end{abstract}


\maketitle

Fabry-Perot cavities, optical resonators, are commonly utilized for
high-precision frequency and distance measurements
\cite{Vaughan:1989}. Currently, kilometer-scale Fabry-Perot cavities
with suspended mirrors are being employed in efforts to detect cosmic
gravitational waves \cite{Abramovici:1992, Bradaschia:1990}. This
application has stimulated renewed interest in cavities with moving
mirrors \cite{Camp:1995, Mizuno:1997, Chickarmane:1998, Pai:2000} and
motivated efforts to model the dynamics of such cavities on the
computer \cite{Redding:1997, Sigg:1997, Bhawal:1998, Beausoleil:1999,
Yamamoto:2000}. Recently, several studies addressed the process of
lock acquisition in which the cavity mirrors move through the
resonance positions \cite{Camp:1995, Lawrence:1999,
Rakhmanov:doppler}. In this process, the Doppler effect due to the
mirror motions impedes constructive interference of light in the
cavity giving rise to complex field dynamics. In contrast, Fabry-Perot
cavities held in the state of resonance have usually been  treated as
essentially static. In this letter, we show that resonant cavities
also have complex field dynamics and we derive the condition for
dynamic resonance. Our analysis is developed for the very long
Fabry-Perot cavities of gravitational wave detectors, but the results
are general and apply to any cavities, especially when the frequencies
of interest are close to the cavity free spectral range.

\begin{figure}
   \centering\includegraphics[width=0.46\textwidth]{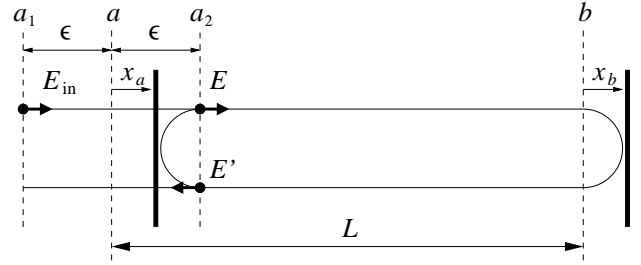}
   \caption{Mirror positions and fields in a Fabry-Perot cavity.}
   \label{fpcav}
\end{figure}

We consider a Fabry-Perot cavity with a laser field incident from
one side as shown in Fig.~\ref{fpcav}. Variations in the cavity length 
are due to the mirror displacements $x_a(t)$ and $x_b(t)$ which are 
measured with respect to reference planes $a$ and $b$. The nominal 
light transit time and the free spectral range (FSR) of the cavity 
are defined by 
\begin{equation}
   T = L/c, \qquad \omega_{\mathrm{fsr}} = \pi/T.
\end{equation}
The field incident upon the cavity and the field circulating in the
cavity are described by plane waves with nominal frequency $\omega$
and wavenumber $k$ ($k=\omega/c$). Variations in the laser frequency 
are denoted by $\delta \omega(t)$. We assume that the mirror
displacements are much less than the nominal cavity length and that 
the deviations of the laser frequency are much less than the nominal
frequency.

At any given place the electric field ${\mathcal{E}}$ in 
the cavity oscillates at a very high frequency: 
${\mathcal{E}}(t) \propto \exp(i\omega t)$. For simplicity, we 
suppress the fast-oscillating factor and define the slowly-varying 
field as $E(t) = {\mathcal{E}}(t) \exp(-i\omega t)$. To properly 
account for the phases of the propagating fields, their complex 
amplitudes are defined at fixed locations, reference planes $a_1$ 
and $a_2$, as shown in Fig.~\ref{fpcav}. (The small offset $\epsilon$
is introduced for convenience and can be set to zero at the end of
calculations.)

The equations for fields in the cavity can be obtained by tracing
a wavefront during its complete round-trip in the cavity (starting
from the reference plane $a_2$). The propagation delays $\tau_1$ and
$\tau_2$ depend on the mirror positions and are given by
\begin{eqnarray}
   c \; \tau_1 & = & L - \epsilon + x_b(t - \tau_1) ,\label{t1} \\
   c \; \tau_2 & = & \epsilon - x_a(t - \tau_2) .\label{t2}
\end{eqnarray}
Then the fields in the cavity satisfy the equations:
\begin{eqnarray}
   E'(t) & = & - r_b E (t - 2 \tau_1) e^{-2i \omega \tau_1} ,
               \label{Epr} \\
   E(t)  & = & - r_a E'(t - 2 \tau_2) e^{-2i \omega \tau_2} +
               \nonumber \\
         &   & t_a E_{\mathrm{in}}(t - 2 \epsilon/c) ,
               \label{Ecav}
\end{eqnarray}
where $r_a$ and $r_b$ are the mirror reflectivities, and $t_a$ is the
transmissivity of the front mirror.

Because the field amplitudes $E$ and $E'$ do not change significantly 
over times of order $x_{a,b}/c$, the small variations in these 
amplitudes during the changes in propagation times due to mirror 
displacements can be neglected. Furthermore, the reference planes $a$ 
and $b$ can be chosen so that the nominal length of the Fabry-Perot 
cavity becomes an integer multiple of the laser wavelength, making 
$\exp(-2ikL)=1$. Finally, the offset $\epsilon$ can be set to zero, 
and Eqs.~(\ref{Epr})-(\ref{Ecav}) can be combined yielding one 
equation for the cavity field
\begin{equation}\label{Eiter}
   E(t) = t_a E_{\mathrm{in}}(t) + 
          r_a r_b E(t - 2T) \exp[-2ik \delta L(t)] .
\end{equation}
Here $\delta L(t)$ is the variation in the cavity length ``seen'' 
by the light circulating in the cavity,
\begin{equation}\label{deltaL2}
   \delta L(t) = x_b(t - T) - x_a(t) .
\end{equation}
Note that the time delay appears in the coordinate of the end mirror, 
but not the front mirror. This is simply a consequence of our
placement of the laser source; the light that enters the cavity 
reflects from the end mirror first and then the front mirror. 
For $\delta L=0$, Laplace transformation of both sides of
Eq.~(\ref{Eiter}) yields the basic cavity response function
\begin{equation}\label{cavTf}
   H(s) \equiv \frac{\tilde{E}(s)}{\tilde{E}_{\mathrm{in}}(s)} = 
      \frac{t_a}{1 - r_a r_b e^{-2sT}},
\end{equation}
where tildes denote Laplace transforms.

The static solution of Eq.~(\ref{Eiter}) is found by considering 
a cavity with fixed length ($\delta L = {\mathrm{const}}$) 
and an input laser field with fixed amplitude and frequency
($A, \delta \omega = {\mathrm{const}}$). In this 
case the input laser field and the cavity field are given by
\begin{eqnarray}
   E_{\mathrm{in}}(t) & = & A   \; e^{i \delta \omega t}, \\
   E(t)               & = & E_0 \; e^{i \delta \omega t},
\end{eqnarray}
where $E_0$ is the amplitude of the cavity field,
\begin{equation}\label{Estat}
   E_0 = \frac{t_a A}{1 - r_a r_b \exp[-2i
      (T \; \delta \omega + k \; \delta L)]} .
\end{equation}
The cavity field is maximized when the length and the laser frequency
are adjusted so that
\begin{equation}\label{statRes}
   \frac{\delta \omega}{\omega} = - \frac{\delta L}{L} .
\end{equation}
This is the well-known static resonance condition.
The maximum amplitude of the cavity field is given by
\begin{equation}\label{barE}
   \bar{E} = \frac{t_a A}{1 - r_a r_b} .
\end{equation}

Light can also resonate in a Fabry-Perot cavity when its length and
the laser frequency are changing. For a fixed amplitude and variable 
phase, the input laser field can be written as
\begin{equation}
    E_{\mathrm{in}}(t) = A \; e^{i \phi(t)} ,
\end{equation}
where $\phi(t)$ is the phase due to frequency variations,
\begin{equation}
   \phi(t) = \int_0^t \delta \omega(t') dt' .
\end{equation}
Then the steady-state solution of Eq.~(\ref{Eiter}) is 
\begin{equation}
    E(t) = \bar{E} \; e^{i \phi(t)} ,
\end{equation}
where the amplitude $\bar{E}$ is given by Eq.~(\ref{barE}) and the 
phase satisfies the condition
\begin{equation}\label{phaseT}
   \phi(t) - \phi(t - 2T) = - 2 k \; \delta L(t) .
\end{equation}
Thus resonance occurs when the phase of the input laser field is 
corrected to compensate for the changes in the cavity length due to
the mirror motions. The associated laser frequency correction is equal 
to the Doppler shift caused by reflection from the moving mirrors
\begin{equation}
   \delta \omega(t) - \delta \omega(t - 2T) = 
      - 2 \frac{v(t)}{c} \omega , 
\end{equation}
where $v(t)$ is the relative mirror velocity ($v = d\delta L/dt$).
The equivalent formula in the Laplace domain is
\begin{equation}\label{dynResF}
   C(s) \frac{\delta \tilde{\omega}(s)}{\omega} = - 
      \frac{\delta \tilde{L}(s)}{L} ,
\end{equation}
where $C(s)$ is the normalized frequency-to-length transfer
function which is given by
\begin{equation}
   C(s) = \frac{1 - e^{-2sT}}{2 s T} .
\end{equation}
Eq.~(\ref{dynResF}) is the condition for dynamic resonance. It must be
satisfied in order for light to resonate in the cavity when the
cavity length and the laser frequency are changing.

The transfer function $C(s)$ has zeros at multiples of the cavity free 
spectral range,
\begin{equation}\label{zeros}
   z_n = i \omega_{\mathrm{fsr}} n ,
\end{equation}
where $n$ is integer, and therefore can be written as the infinite
product,
\begin{equation}
   C(s) = e^{-s T} \prod \limits_{n=1}^{\infty} \left( 
      1 - \frac{s^2}{z_n^2} \right) ,
\end{equation}
which is useful for control system design\footnote{This formula is 
derived using the infinite-product representation for sine:
$\sin x = x \;\prod_{n=1}^{\infty}\left(1 - x^2/\pi^2 n^2\right)$.}.

To maintain resonance, changes in the cavity length must be
compensated by changes in the laser frequency according to 
Eq.~(\ref{dynResF}). If the frequency of such changes is much 
less than the cavity free spectral range, $C(s) \approx 1$ and 
Eq.~(\ref{dynResF}) reduces to the quasi-static approximation,
\begin{equation}\label{qstatResF}
   \frac{\delta \tilde{\omega}(s)}{\omega} \approx - 
      \frac{\delta \tilde{L}(s)}{L} .
\end{equation}
At frequencies above the cavity free spectral range, 
$C(s) \propto 1/s$
and increasingly larger laser frequency changes are required to
compensate for cavity length variations. Moreover, at multiples of
the FSR, $C(s)=0$ and no frequency-to-length compensation is possible.

In practice, Fabry-Perot cavities always deviate from resonance, and 
a negative-feedback control system is employed to reduce the
deviations. For small deviations from resonance, the cavity field can
be described as
\begin{equation}
   E(t) = [\bar{E} + \delta E(t)] e^{i\phi(t)},
\end{equation}
where $\bar{E}$ is the maximum field given by Eq.(\ref{barE}), and 
$\delta E$ is a small perturbation ($|\delta E| \ll |\bar{E}|$). 
Substituting this equation into Eq.~(\ref{Eiter}), we see that the 
perturbation evolves in time according to
\begin{eqnarray}
   & & \delta E(t) - r_a r_b \delta E(t - 2T) = \nonumber \\
   & & \qquad - i r_a r_b \bar{E} \left[ \phi(t) - \phi(t - 2T) + 
       2 k \; \delta L(t) \right] .\label{dE(t)}
\end{eqnarray}
This equation is easily solved in the Laplace domain, yielding
\begin{equation}\label{dE(s)}
   \delta \tilde{E}(s) = - i r_a r_b \bar{E} \; 
      \frac{ \left(1 - e^{-2sT} \right) \tilde{\phi}(s) +
      2 k \; \delta \tilde{L}(s)}{1 - r_a r_b e^{-2sT}} .
\end{equation}

\begin{figure}
   \centering\includegraphics[width=0.46\textwidth]{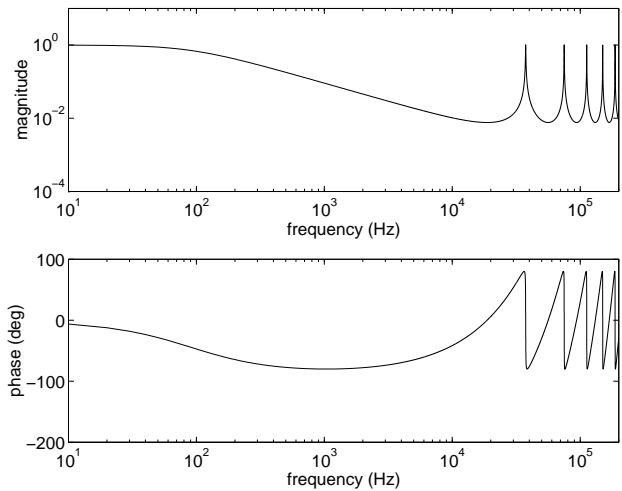}
   \caption{Bode plot of $H_L(i\Omega)$ for the LIGO 4-km
   Fabry-Perot cavities. The peaks occur at multiples of the 
   FSR ($37.5$ kHz) and their half-width ($91$ Hz) is equal to the 
   inverse of the cavity storage time.}
   \label{tfLength}
\end{figure}

Deviations of the cavity field from its maximum value can be measured
by the Pound-Drever-Hall (PDH) error signal which is widely utilized
for feedback control of Fabry-Perot cavities \cite{Drever:1983}. The
PDH signal is obtained by coherent detection of phase-modulated light
reflected by the cavity. With the appropriate choice of the
demodulation phase, the PDH signal is proportional to the imaginary
part of the cavity field (Eq.~(\ref{dE(s)})) and therefore can be 
written as
\begin{equation}\label{PDH}
   \delta \tilde{V}(s) \propto H(s) \left[
      \frac{\delta \tilde{L}(s)}{L} + C(s) \;
      \frac{\delta \tilde{\omega}(s)}{\omega} \right] ,
\end{equation}
where $H(s)$ is given by Eq.~(\ref{cavTf}). In the presence of length
and frequency variations, feedback control will drive the error signal 
toward the null point, $\delta \tilde{V} = 0$, thus maintaining
dynamic resonance according to Eq.~(\ref{dynResF}).

The response of the PDH signal to either length or laser frequency 
deviations can be found from Eq.~(\ref{PDH}). The normalized
length-to-signal transfer function is given by
\begin{equation}
   H_L(s) = \frac{H(s)}{H(0)} = 
      \frac{1 - r_a r_b}{1 - r_a r_b e^{-2 s T}}.
\end{equation}
A Bode plot (magnitude and phase) of $H_L$ is shown in 
Fig.~\ref{tfLength} for the LIGO \cite{Abramovici:1992} 
Fabry-Perot cavities with $L=4$ km, $r_a=0.985$, and $r_b=1$.
The magnitude of the transfer function,
\begin{equation}\label{airy}
   |H_L(i\Omega)| = \frac{1}{\sqrt{1 + F \sin^2 \Omega T}},
\end{equation}
is the square-root of the well-known Airy function with the 
coefficient of finesse $F = 4 r_a r_b/(1 - r_a r_b)^2$. 
(In optics, the Airy function describes the intensity profile 
of a Fabry-Perot cavity \cite{Born:1980}.)

\begin{figure}
   \centering\includegraphics[width=0.46\textwidth]{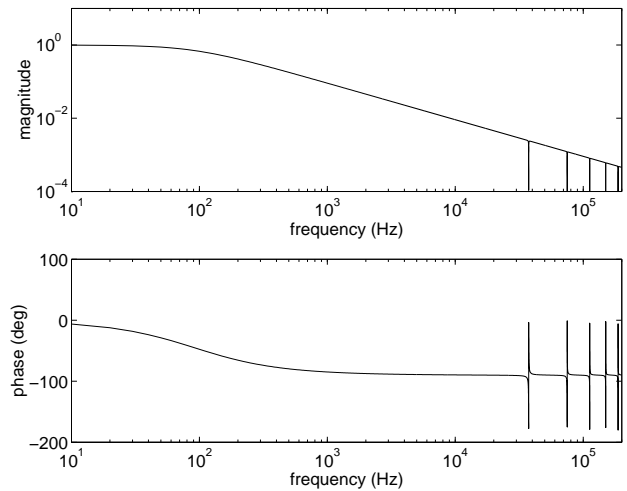}
   \caption{Bode plot of $H_{\omega}(i\Omega)$. The sharp features 
   are due to the zero-pole pairs at multiples of the FSR.}
   \label{tfFreq}
\end{figure}

The transfer function $H_L$ has an infinite number of poles:
\begin{equation}\label{poles}
   p_n = - \frac{1}{\tau} + i \omega_{\mathrm{fsr}} n ,
\end{equation}
where $n$ is integer, and $\tau$ is the storage time of the cavity,
\begin{equation}
   \tau = \frac{2T}{|\ln (r_a r_b)|}.
\end{equation}
Therefore, $H_L$ can be written as the infinite product,
\begin{equation}
   H_L(s) = e^{s T} \prod \limits_{n=-\infty}^{\infty}
      \frac{p_n}{p_n - s},
\end{equation}
which can be truncated to a finite number of terms for use in 
control system design.

The response of a Fabry-Perot cavity to laser frequency variations 
is very different from its response to length variations. The
normalized frequency-to-signal transfer function is given by 
\begin{equation}
   H_{\omega}(s) = C(s) H_L(s),
\end{equation}
or, more explicitly as
\begin{equation}
   H_{\omega}(s) = \left( \frac{1 - e^{-2 s T}}{2 s T} \right) \;
      \left( \frac{1 - r_a r_b}{1 - r_a r_b e^{-2 s T}} \right).
\end{equation}
A Bode plot of $H_{\omega}$, calculated for the same parameters as 
for $H_L$, is shown in Fig.~\ref{tfFreq}. The transfer function 
$H_{\omega}$ has zeros given by Eq.~(\ref{zeros}) with $n \neq 0$,
and poles given by Eq.~(\ref{poles}). The poles and zeros come in
pairs except for the lowest order pole, $p_0$, which does not have 
a matching zero. Therefore, $H_{\omega}$ can be written as the
infinite product,
\begin{equation}
   H_{\omega}(s) = \frac{p_0}{p_0 - s}
      {\prod\limits_{n = -\infty}^{\infty}}'
      \left( \frac{1 - s/z_n}{1 - s/p_n} \right) ,
\end{equation}
where the prime indicates that $n=0$ term is omitted from the
product.

The zeros in the transfer function indicate that the cavity does not 
respond ($\delta E = 0$) to laser frequency deviations if these 
deviations occur at multiples of the cavity FSR. In this case, 
the amplitude of the circulating field is constant while the 
phase of the circulating field is changing with the phase of the 
input laser field.

In summary, we have shown that resonance can be maintained in a
Fabry-Perot cavity even when the cavity length and laser frequency
are changing. In this dynamic resonance state, changes in the laser
frequency and changes in  the cavity length play very different roles
(Eq.~(\ref{dynResF})) in contrast to the quasi-static resonance state
where they appear equally (Eq.~(\ref{qstatResF})). Maintenance of
dynamic resonance requires that the  frequency-to-length transfer
function, $C(s)$, be taken into account when compensating for length
variations by frequency changes and vice versa. Compensation for
length variations by frequency changes becomes increasingly more
difficult at frequencies above the FSR, and impossible at multiples
of the FSR.

As can be seen in Fig.~\ref{tfFreq}, the response of the PDH error 
signal to laser frequency variations decreases as $1/\Omega$ for
$\Omega\gg\tau^{-1}$ and becomes strongly suppressed at frequencies 
equal to multiples of the cavity FSR. In contrast, the response  
to length variations is a periodic function of frequency as shown in 
Fig.~\ref{tfLength}. For $\Omega\gg\tau^{-1}$, it also decreases as
$1/\Omega$ but only to the level of $(1+F)^{-\frac{1}{2}}$ and then 
returns to its maximum value. Thus, at multiples of the FSR, the 
sensitivity to length variations is maximum while the sensitivity to 
frequency variations is minimum.

Both these features  suggest searches for gravitational waves at
frequencies near multiples of the FSR. However, because gravitational
waves interact with the light as well as the mirrors, the response of
an optimally-oriented interferometer is equivalent to $H_{\omega}(s)$
and not to $H_L(s)$ \cite{Mizuno:1997}. Thus, an optimally-oriented
interferometer does not respond to gravitational wave at multiples of
the FSR. However, for other orientations gravitational waves can be
detected with enhanced sensitivity at multiples of the cavity FSR
\cite{Schilling:1997}.

We thank Robert Coldwell, Guido Mueller and David Shoemaker for 
comments on the paper. This research was supported by the National 
Science Foundation under grants PHY-9210038 and PHY-0070854.

\end{document}